\documentclass[a4paper,12pt]{article}

\pdfoutput=1 

\usepackage{jcappub} 

\usepackage[T1]{fontenc} 
\usepackage{comment}
\usepackage[top=1in,bottom=1in,left=1in,right=1in]{geometry}

\usepackage{times}

\newcommand{\elena}[1]{{\color{blue}(emassara@flatironinstitute.org)}}

\begin{document}

\flushbottom
\thispagestyle{empty}
\pagestyle{plain}

\begin{center}
\LARGE{Astro2020 Science White Paper}\\    
\vspace{40pt}
\Huge{Cosmic voids: a novel probe }\\
\Huge{ to shed light on our Universe}
\end{center}
\normalsize{}
\vspace{30pt}
\normalsize{}
\noindent{\textbf{Thematic areas:}}\hspace{16pt}Primary: 7. Cosmology and Fundamental Physics

\hspace{70pt} Secondary: 6. Galaxy Evolution
\vspace{5pt}

\vspace{20pt}

\noindent \textbf{Main authors:} 
\hspace{20pt}Alice Pisani\textsuperscript{a,1} (Princeton)\note{apisani@astro.princeton.edu}

\hspace{70pt} Elena Massara\textsuperscript{b,2} (CCA)\note{emassara@flatironinstitute.org}

\hspace{70pt} David N. Spergel\textsuperscript{a,b} (CCA/Princeton)

\vspace{20pt}

\noindent\textbf{Co-authors:} 
David Alonso\textsuperscript{c} (U. of Oxford), 
Tessa Baker\textsuperscript{d} (Queen Mary U. London), 
Yan-Chuan Cai\textsuperscript{e} (U. of Edinburgh), 
Marius Cautun\textsuperscript{f} (Durham U.), 
Christopher Davies\textsuperscript{f} (Durham U.), 
Vasiliy Demchenko\textsuperscript{e} (U. of Oxford), 
Olivier Dor\'e\textsuperscript{g,h} (JPL/Caltech),
Andy Goulding\textsuperscript{a} (Princeton),
M\'elanie Habouzit\textsuperscript{b} (CCA),
Nico Hamaus\textsuperscript{i} (LMU), 
Adam Hawken\textsuperscript{j} (CPPM), 
Christopher M. Hirata\textsuperscript{k,l,m} (Ohio State U.), 
Shirley Ho\textsuperscript{a,b} (CCA/Princeton), 
Bhuvnesh Jain\textsuperscript{n} (UPenn), 
Christina D. Kreisch\textsuperscript{a} (Princeton), 
Federico Marulli\textsuperscript{o,p,q} (U. of Bologna/INAF), 
Nelson Padilla\textsuperscript{r,s} (UC), 
Giorgia Pollina\textsuperscript{i} (LMU), 
Martin Sahl\'en\textsuperscript{t} (Uppsala U.), 
Ravi K. Sheth\textsuperscript{n,u} (UPenn/ ICTP), 
Rachel Somerville\textsuperscript{b} (CCA), 
Istvan Szapudi\textsuperscript{v} (U. of Hawaii), 
Rien van de Weygaert\textsuperscript{w} (U. of Groningen), 
Francisco Villaescusa-Navarro\textsuperscript{b} (CCA), 
Benjamin D. Wandelt\textsuperscript{a,b,x,y} (IAP /ILP/CCA/ Princeton),
Yun Wang\textsuperscript{z,$\alpha$} (Caltech/IPAC).

\vspace{3pt}
\noindent \textit{(Authors complete affiliations can be found after the references.)}

\vspace{20pt}
\noindent{\textbf{Abstract:}}\\Cosmic voids, the less dense patches of the Universe, are promising laboratories to extract cosmological information. Thanks to their unique low density character, voids are extremely sensitive to diffuse components such as neutrinos and dark energy, and represent ideal environments to study modifications of gravity, where the effects of such modifications are expected to be more prominent. Robust void-related observables, including for example redshift-space distortions (RSD) and weak lensing around voids, are a promising way to chase and test new physics. Cosmological analysis of the large-scale structure of the Universe predominantly relies on the high density regions. Current and upcoming surveys are designed to optimize the extraction of cosmological information from these zones, but leave voids under-exploited.
A dense, large area spectroscopic survey with imaging capabilities is ideal to exploit the power of voids fully. Besides helping illuminate the nature of dark energy, modified gravity, and neutrinos, this survey will give access to a detailed map of under-dense regions, providing an unprecedented opportunity to observe and study a so far under-explored galaxy population. \\

\newpage
\pagenumbering{gobble} 
\pagenumbering{arabic}
\normalsize{}
\section{\normalsize{Introduction}}
\label{sec:intro}
\vspace{-5pt}
Cosmic voids, the large under-dense regions in the distribution of galaxies \cite{Gregory_1978,Zeldovich_1982,Einasto_1980,Chincarini_1975,Kirshner_1981}, are a powerful and not yet fully utilized  tool for studying cosmology and galaxy formation.
Together with halos (high density regions which have undergone virialization), filaments, and walls, voids constitute the large-scale structure of the Universe, known as the cosmic web \cite{Bond_1996,Bond_1996b}.
Void sizes span from tens to hundreds of Mpc; they fill most of the volume of the Universe.

The large-scale structure of the Universe, the Cosmic Microwave Background (CMB) and Supernovae are powerful probes to extract robust cosmological information and have been widely used to increase our understanding of cosmology \cite{PlanckCollaboration_2018}. Nevertheless many unknowns remain. Our Universe is undergoing an accelerated expansion, but the origin of such acceleration is still not understood.
The standard model of cosmology invokes the presence of a new component, dubbed dark energy, that would constitute $\simeq$ 70\% of our Universe, but the nature and behaviour of this component remain mysterious. Additionally, while we know that about 26\% of the matter-energy content of the Universe is made of dark matter, we can only measure it through its gravitational effects. To shed light on the nature of dark matter and dark energy it is important to employ novel and orthogonal methods.

Traditionally, studies of the galaxy distribution at large scales rely on galaxy two-point statistics and baryonic acoustic oscillations (BAO) \citep{Eisenstein_2005} imprinted on it. Since the galaxy distribution is non-Gaussian at late times, additional information is contained in higher order statistics. 
Moreover, in the Gaussian initial conditions the cosmological volume is split between high and low-density regions, each kind of density region containing part of the cosmological information. 
Standard galaxy clustering analysis---even when it considers higher order tools (e.g. bispectrum, trispectrum)---would only be sensitive to collapsed regions corresponding to positive density fluctuations. In dense regions virialization erases some memory of initial conditions, conversely low-density regions remain much more linear and keep such memory.

Cosmic voids provide access to both the high order information \cite{White_1979,Hamilton_1985,Fry_1985} and the information on initial conditions. This makes them particularly sensitive to new physics and hence optimal laboratories to gain insights into our Universe.

Current surveys (e.g., eBOSS \cite{eBOSS}) cover large regions of the sky and extend over large redshift ranges. They already provide us with a mapping of the under-dense regions, containing cosmological information at scales between $10-100\,h^{-1}\mathrm{Mpc}$. We are now able to robustly identify these under-dense regions with a variety of algorithms \citep{Colberg_2008} using both spectroscopic \citep{Hamaus_2016,Mao_2016, Sutter_2012,Sutter_2014,Micheletti_2014} and photometric data \citep{Sanchez_2017,Pollina_2018}, for which the accuracy is constantly improving.

In this paper we describe in Section \ref{sec:constraints} the potential of voids to provide us with new insights for cosmology and galaxy formation and in Section
\ref{sec:challenges} the challenges for the coming decade. 
Section \ref{sec:survey} discusses surveys needs for voids in the framework of the Astro2020 Decadal survey and emphasizes the opportunities of WFIRST and ground-based assets for void science.

\section{\normalsize{Understanding the Universe with cosmic voids}}
\label{sec:constraints}
\vspace{-5pt}
Our knowledge of the Universe is far to be complete. Measurements of the statistical properties of voids can fill this gap: they are sensitive to structure growth, dark energy, modified gravity, sum of neutrino masses and galaxy formation.

\vspace{-3pt}
\subsection{How do structures grow?}
\label{sec:growthrate}
\vspace{-5pt}

The peculiar velocities of galaxies leave a characteristic pattern in their observed spatial distribution, known as redshift-space distortions (RSDs). RSDs are sensitive to the growth rate of structures, $f(z)=\mathrm{d}\ln D / \mathrm{d}\ln a$, the derivative of the amplitude of linear density fluctuations with respect to the scale factor~\cite{Peacock_2001,Guzzo_2008}. The growth rate depends on the matter content of the Universe and its measurement allows to test General Relativity~\cite{ Acquaviva_2008}. While RSDs have already been analyzed extensively in the two-point statistics of galaxies, a major challenge is posed by the complexity of their non-linearity on small scales (below $\sim 20h^{-1}\mathrm{Mpc}$), where perturbative methods fail and the available cosmological information is fundamentally limited by the process of shell crossing and virialization. One way to circumvent this limitation is to focus on non-virialized regions of the Universe: cosmic voids. Voids are dominated by coherent single-stream motions~\cite{Shandarin_2011}. This makes their dynamical description much simpler than any other type of structure. RSDs reveal this dynamical information via the anisotropy of the void-galaxy cross-correlation function in redshift space~\cite{Hamaus_2015,Cai_2016}. The latter can be directly linked to the growth rate $f(z)$, either via the Gaussian streaming model \cite{Hamaus_2016, Achitouv_2017, Hawken_2016}, or a multipole decomposition (e.g. \cite{Cai_2016,Hamaus_2017}). At the same time, an accurate model for void RSDs ensures unbiased cosmological constraints from geometric distortions via the Alcock-Paczy\'nski test~\cite{Hamaus_2016,Sutter_2012,Hamaus_2014,Mao_2016} (see Section \ref{sec:darkenergy}). Up to now the RSD analysis using voids from the SDSS BOSS data provided the tightest independent constraints from void data alone on $\Omega_m$ (10\% accuracy) and on $f/b$ (12\% accuracy, $b$ is the galaxy bias) \cite{Hamaus_2016,Hamaus_2017}. The accuracy on $f/b$ from voids is hence already competitive with state-of-the-art galaxy clustering constraints (cf. \cite{Chuang_2017} and \cite{Hamaus_2017}), while the AP test constraining $\Omega_m$ with voids outperforms the latter by a factor of $\sim 4$ (cf. \cite{GilMarin2016} and \cite{Hamaus_2016}, with 1\% accuracy on void shape Alcock-Paczy\'nski measurement).
\vspace{-3pt}
\subsection{What is dark energy?}
\label{sec:darkenergy}
\vspace{-5pt}
The accelerated expansion of the Universe is one of the most puzzling discoveries of the 20th century.  Large missions in the 2020s (DESI\cite{DESI}, SPHEREx\cite{SPHEREx}, Euclid\cite{Euclid}, LSST \cite{LSST}, WFIRST\cite{WFIRST, Dore_2019}), aim to study the nature of dark energy. Many of these experiments have been optimized to use two-point statistics (weak gravitational lensing or BAO) to probe  the possibility of a time-evolving equation of state for dark energy, often parametrized as $w(z)=w_0+ w_a[z/(z+1)]$ \cite{Chevallier_2001,Linder_2003}. Cosmic voids, devoid of matter by definition, are \textit{dark energy-dominated} objects: their study is a novel route to explore dark energy. The evolution of voids is ruled by the joint action of gravitational attraction, that empties voids by pushing material towards their boundaries \cite{Icke_1984,vandeweygaert_1993, Cautun_2016}, and the expansion of the Universe, that also enlarges voids by diluting the space between galaxies. Throughout cosmic history, the expansion of voids differs depending on whether dark energy is constant, or evolving with time. This gives rise to a dependence of void size on dark energy: the void size function (the number of voids as a function of radius \cite{Platen_2008}) is highly sensitive to the dark energy equation of state \cite{Bos_2012, Pisani_2015}. In addition, when converting the angles and redshifts of observed celestial objects (RA,Dec,$z$) into distances, we need to assume a fiducial cosmology. In a homogeneous and isotropic universe their average shape is expected to be spherical \cite{Lavaux_2012} (even if voids have highly non-spherical shapes individually). If we are using the correct model to interpret data, we will observe an average void shape that is indeed spherical---if not, our cosmological model is wrong. This method to validate a model, known as the Alcock-Paczy\'nski (AP) test, can be applied to void stacks and is particularly sensitive to the dark energy equation of state (potentially outperforming traditional galaxy clustering constraints on $w_0$ and $w_a$ \cite{Lavaux_2012}---provided that RDSs are modelled).

\vspace{-3pt}
\subsection{Should we modify the laws of gravity?}
\vspace{-5pt}
Extensions of Einstein's theory of General Relativity (GR) are of intense interest at present, as potential explanations of cosmic acceleration \citep{Clifton_2012,Ishak_2019}. Features of these theories include ``screening mechanisms'', that allow to recover GR near astrophysical objects \citep{Joyce_2015}. Conversely, in low-density void environments, the new gravitational fields introduced by most modified gravity theories (e.g. the Horndeski scalar field \cite{Horndeski_1974}) are fully dynamical. It is in cosmic voids where these alternative theories deviate most strongly from GR. 
Observationally, modified gravity leads to a few consequences. Firstly, it leads to a difference in the matter density profile compared with GR that can be measured via gravitational lensing around voids \cite{Spolyar_2013,Melchior_2014,Clampitt_2015, Cai_2015, Barreira_2015,Sanchez_2017, Baker_2018,Davies_2018}. This observable is particularly powerful for models with strong deviation between the lensing potential and the Newtonian potential, since a significantly enhanced lensing signal can arise compared to GR, even for the same density profile \cite{Baker_2018}. Secondly, modified gravity causes a faster expansion of voids that can be captured by RSDs around voids (see e.g. \cite{Hamaus_2015,Hamaus_2016,Hamaus_2017,Cai_2016, Hawken_2016, Achitouv_2017,Falck_2018, Nadathur_2019a, Nadathur_2019b} and Section \ref{sec:growthrate}). Finally, modified gravity results in environmental differences that can be observed by comparing astrophysical features such as galaxy properties in over-dense and void regions (e.g. \cite{Hui_2009, Jain_2011a, Jain_2011b, Zhao_2010, Cabre_2012}).

\vspace{-3pt}
\subsection{What is the mass of neutrinos?}
\vspace{-5pt}
Neutrinos remain one of the most elusive components of the standard model. The discovery of neutrino oscillations proved that neutrinos must have mass \cite{Becker-Szendy_1992,Fukuda_1998,Ahmed_2004}, whereas the standard model of particle physics assumes they are massless. Cosmology places some of the strongest upper bounds on the sum of neutrino masses (e.g. $\Sigma m_\nu < 0.12\,\mathrm{eV}$ at 95\% CL for Planck TT,TE,EE + low E + lensing + BAO \cite{PlanckCollaboration_2018}). Such constraints approach the sensitivity needed to determine the neutrino mass hierarchy, since we expect $\Sigma m_\nu > 0.1$ eV for the inverted mass hierarchy \cite{PlanckCollaboration_2018}. Neutrinos cluster like cold dark matter (CDM) and contribute to structure formation on scales larger than their free-streaming length, which depends on the neutrino species' mass. For smaller scales, neutrinos suppress structure formation and this effect increases with their mass \cite{Lesgourgues_2006}. Typical sizes of voids span the range of neutrino free-streaming scales. Additionally, cosmic voids, being devoid of matter, are particularly sensitive to neutrinos (and all diffuse components) since the mass fraction of neutrinos with respect to CDM is higher in voids than in high density regions.
 For these reasons, voids are becoming a topic of increasing interest for studying neutrinos \cite{Villaescusa_2013,Massara_2015,Banerjee_2016,Kreisch_2018,Dvorkin_2019}.
Void-related observables, such as number, size, shape, distribution and clustering of cosmic voids, are powerful probes of neutrino properties. Recent results show that the galaxy bias plays an important role in the effect of neutrinos on void observables \cite{Kreisch_2018}, by changing how neutrino masses impact void clustering. A theoretical model for void properties in neutrino cosmology is necessary to exploit the constraining power of future surveys.
 
\vspace{-3pt}
\subsection{How do galaxies evolve?}
\vspace{-5pt}
Galaxy evolution is the complex result of {\it nature} and {\it nurture}, i.e. of internal processes and external effects, such as large-scale environment (e.g. stripping and harassment, particularly effective in dense regions) and galaxy mergers. While disentangling these contributions in high-density environments is very challenging, galaxies in cosmic voids represent a unique laboratory containing isolated systems whose evolution has been driven almost completely by in-situ processes. In the hierarchical model of structure formation, which assumes that small galaxies progressively merge to form larger galaxies with time, galaxies embedded in cosmic voids would represent less evolved systems. Void galaxies (e.g. observed in SDSS) have properties significantly different from similar objects in denser environments; they tend to have stellar disks with smaller radii \citep{2011ASSP...27...17V}, to be star-forming blue galaxies \citep{1999AJ....118.2561G, 2004ApJ...617...50R}, with later-type morphological types \citep{2011ASSP...27...17V,2014MNRAS.445.4045R, Beygu_2017}, i.e. more spiral and less elliptical galaxies at a given stellar mass or absolute magnitude. Most void galaxies are still on the main galaxy star-forming sequence (i.e. forming stars efficiently), which seems to favor either a slow or a late galaxy evolution model.
Current observational studies of void galaxies do not always reach a consensus (mainly due to small statistics). This will be improved by upcoming surveys (e.g. PFS \cite{PFS}, WFIRST, Euclid) and the advent of large volume state-of-the-art cosmological hydrodynamical simulations. It will allow us to make self-consistent comparisons between simulated and observed void galaxies in order to better understand the role of galaxy mergers and environment on galaxy evolution.
\vspace{-5pt}
\section{\normalsize{Challenges for the coming decade}}
\label{sec:challenges}
\vspace{-5pt}
Section \ref{sec:constraints} presented the main current unknowns in our cosmological panorama, and argued that cosmic voids are an ideal laboratory to provide answers to these unknowns. In this Section we discuss the challenges to face in order to optimally exploit voids. 
\vspace{-3pt}
\subsection{Simulation challenges}
\label{sec:sims}
\vspace{-5pt}
When measuring cosmological parameters with void-related observables we need to understand their degeneracies. For example, the void size function is affected by both properties of dark energy and neutrinos. RSDs are sensitive to modifications of gravity and neutrinos. Furthermore, the anisotropies measured by the AP test can be due to both dark energy and modifications of gravity. These degeneracies add up with other well-known degeneracies in cosmology (such as $\Omega_m$--$\sigma_8$). 
One possibility to reduce degeneracies between cosmological parameters is to combine void-related observables with other probes (e.g. CMB, clusters, galaxy power spectrum \cite{Sahlen_2015,Sahlen_2016,Sahlen_2018}). To fully exploit the potential of voids it is necessary to understand these degeneracies with simulations dedicated to the study of cosmic voids. Since voids are the largest elements of the cosmic web, they are the biased tracers of the underlying matter density field that have the lowest number density. For this reason, large cosmological volumes are needed to get sufficient statistics that mitigate sample/cosmic variance. On the other hand, both the large size of voids and the fact that they are pristine environments make them less sensitive to baryonic effects such as galaxy feedback~\citep{Paillas_2016}. Hence, the properties of cosmic voids can be studied with N-body dark matter simulations instead of full hydrodynamical simulations. 

While it is useful to study voids in the dark matter distribution to allow comparisons with theoretical models, voids identified in the halo distribution are more directly connected to observations. 
The required mass resolution of the N-body simulations will depend on the tracer used to identify voids: a low-resolution to find voids in the matter field, and a high-resolution to search for voids in low-mass halo density field. The ultimate goal being to study voids identified in data, that is in the galaxy distribution, the optimal choice is to mimic the galaxy density field. The fastest method to do so is to populate halos with galaxies with the Halo Occupation Distribution (HOD) \cite{Zehavi_2004} framework. Although the use of HOD is generally accepted, we must remember that it is built to reproduce the number density and the two-point correlation function of an observed galaxy population, but higher order statistics might be inaccurate. Further studies will be necessary in upcoming years to test if voids identified in HOD-constructed galaxy fields can provide reliable catalogs. Finally, an important ingredient in traditional Bayesian analysis is the need for a covariance matrix. To compute it accurately, hundreds to thousands of simulations are necessary. Algorithms to approximate N-body simulations, such as COLA~\cite{Tassev_2013} or FastPM~\cite{Feng_2016}, can be used to build these simulations.

\vspace{-3pt}
\subsection{Theoretical challenges}
\vspace{-5pt}
Together with simulations, a major challenge for void science is to further develop the theoretical framework that describes cosmic voids. In particular, it is important to be able to model the type of voids we identify in simulations and in the real data from galaxy surveys. We need to improve the predictions for the void size function \cite{Sheth_2003,Paranjape_2011,Jennings_2013, Voivodic_2016,Pisani_2015,Ronconi_2019}, the density and velocity profiles around voids \cite{Hamaus_2014_universal,Cautun_2016,Demchenko_2016, Massara_2018}, and the two-point statistics of voids \cite{Hamaus_2013,Chan_2014,Cai_2016}. 
While voids in the matter distribution offer the most direct link to theoretical models, observationally we mainly detect voids in the distribution of galaxies (aside from the recent possibility to detect voids with weak lensing maps \cite{Davies_2018}). Therefore, we need to develop models (and mock catalogs, see Section \ref{sec:sims}) that fully
account for the galaxy bias and its impact on void properties \cite{Sutter_2013,Pollina_2017}. Moreover observations are in redshift space, so we will need to further develop tools to take this into account when identifying voids and describing their structure in redshift space. Also, the sub-structure inside voids bears relevant cosmological information, but we currently lack of a model describing it. Furthermore, a robust understanding of the void-in-cloud process---affecting the small under-densities embedded in over-dense regions---is necessary. Finally, these models will need to be extended to the case of non-standard cosmological models.
\vspace{-10pt}
\section{\normalsize{Optimizing surveys for void science}}
\label{sec:survey}
\vspace{-5pt}

When building a survey a compromise is necessary between how deep and how wide it should be---leading to a trade-off between a high tracer density, to resolve the cosmic web in great detail, and a large volume, to access large scales and improve statistics.

Void science needs both \textit{volume and high tracer density}, which is one of the reasons why void observations have not reached their full potential as of today. Indeed a higher density of galaxies will provide access to a detailed mapping of sub-structure in the under-dense regions---as well as unlock the power of different science cases, see Astro2020 Science White Paper by Wang et al. (2019) \cite{Wang_2019}. For void science, a large volume is also necessary to guarantee small error bars, since voids are much sparser than galaxies or even clusters. Voids have so far been studied on surveys optimized for extracting the BAO signal ($100~\,h^{-1}\mathrm{Mpc}$ scale) and measuring RSDs from the galaxy correlation function. 

Fortunately, upcoming surveys will improve the landscape for void science, by observing a larger volume. These surveys include: DESI (2019; 14,000 sq. deg.), SPHEREx (2023; 41,000 sq. deg.); and Euclid (2022; 15,000 sq. deg.). By 2025 completed observations or first data releases from these programs will be available, providing large statistics for voids. 

At this stage, the next natural step to reach exciting gains for void science can be obtained by increasing the galaxy number density and maintaining a large observational volume: the optimal strategy would likely be a wide and dense survey. 
WFIRST, a large mission designed to observe a high tracer density (2026; nominally 2,200 sq. deg.) will follow the aforementioned surveys and has the capability to adjust its survey strategy in response to earlier results. In this framework the optimal strategy would be to consider extensions of the WFIRST high redshift spectroscopic program to be optimized for voids. For this purpose, more of its observing time would need to be devoted to spectroscopy. This will allow to exploit the power of WFIRST's large collecting area and to carry out a survey that is both wide (up to 10,000 sq. deg.) and deep. These observations could be complemented by DESI, PFS, SPHEREx or another survey designed to go to high number density at lower redshift ($z<1$). Furthermore, the imaging capabilities coupled with high density of the WFIRST program will be a strong asset, allowing to perform lensing measurements around voids. Also, the combination with CMB measurements will allow the study of the integrated Sachs-Wolfe effect on voids' locations \cite{Granett_2008, Flender_2013, Cai_2014, Nadathur_2016, Cai_2017, Kovacs_2017}.

The next decade will be a golden age for voids by fully realizing the potential of our void science program.

\vspace{-5pt}
\section{\normalsize{Conclusion}}
\vspace{-10pt}
In this paper we presented the case for void science, arguing that cosmic voids are a novel probe to constrain modified gravity, dark energy, the sum of neutrino masses and galaxy evolution. Voids will answer some of the most relevant questions in cosmology and astrophysics over the next decade. 
We identified theoretical and simulation advances necessary to reach this goal. The large and dense survey we consider will be ideal to exploit void science at its best.


\bibliographystyle{JHEP}
\bibliography{Bibliography} 

\vspace{30pt}

\noindent\textbf{Authors affiliations:}\\
\small
\noindent
\textsuperscript{a}{Department of Astrophysical Sciences, Princeton University, Peyton Hall, 4 Ivy Lane, Princeton, NJ 08544, USA}\\
\textsuperscript{b}{Center for Computational Astrophysics, Flatiron Institute, 162 5th Avenue, New York, NY 10010 USA}\\
\textsuperscript{c}{Department of Physics, University of Oxford, Denys Wilkinson Building, Keble Road, Oxford OX1 3RH, United Kingdom}\\
\textsuperscript{d}{School of Physics and Astronomy, Queen Mary University of London, Mile End Road, London E14NS}\\
\textsuperscript{e}{Institute for Astronomy, University of Edinburgh, Royal Observatory, Edinburgh EH9 3HJ, UK}\\
\textsuperscript{f}{Institute of Computational Cosmology, Department of Physics, Durham University, South Road, Durham, DH1 3LE, UK}\\
\textsuperscript{g}{Jet Propulsion Laboratory, California Institute of Technology, 4800 Oak Grove Dr., Pasadena, CA 91109, USA}\\
\textsuperscript{h}{Space Sciences Laboratory and Theoretical Astrophysics Center, University of California, Berkeley, CA 94720, USA}\\
\textsuperscript{i}{Universit\"{a}ts-Sternwarte M\"{u}nchen, Fakult\"{a}t f\"{u}r Physik,
Ludwig-Maximilians Universit\"{a}t, Scheinerstr. 1, 81679 M\"{u}nchen, Germany}\\
\textsuperscript{j}{CPPM, Aix Marseille University, 163, avenue de Luminy, Marseille, France}\\
\textsuperscript{k}{Center of Cosmology and Astroparticle Physics, The Ohio State University, 191 West Woodruff Ln., Columbus, Ohio 43210, USA},\\ \textsuperscript{l}{Department of Physics, The Ohio State University, 191 West Woodruff Ln., Columbus OH 43210 USA},\\
\textsuperscript{m}{Department of Astronomy, The Ohio State University, 140 West 18th Av., Columbus OH 43210 USA},\\
\textsuperscript{n}{Department of Physics and Astronomy, University of Pennsylvania, Philadelphia, PA 19104, USA}\\
\textsuperscript{o}{Dipartimento di Fisica e Astronomia - Alma Mater Studiorum Universit\`{a} di Bologna, via Piero Gobetti 93/2, I-40129 Bologna, Italy}\\
\textsuperscript{p}{INAF - Osservatorio di Astrofisica e Scienza dello Spazio di Bologna, via Piero Gobetti 93/3, I-40129 Bologna, Italy}\\
\textsuperscript{q}{INFN - Sezione di Bologna, viale Berti Pichat 6/2, I-40127 Bologna, Italy}\\
\textsuperscript{r}{Instituto de Astrof\'{i}sica, Pontificia Universidad Cat\'{o}lica de Chile, Santiago, Chile}\\
\textsuperscript{s}{Centro de Astro-Ingenier\'{i}a, Pontificia Universidad Cat\'{o}lica de Chile, Santiago, Chile}\\
\textsuperscript{t}{Department of Physics and Astronomy, Uppsala University, SE-751 20 Uppsala, Sweden}\\
\textsuperscript{u}{The Abdus Salam International Center for Theoretical Physics, Strada Costiera, 11, Trieste 34151, Italy}\\
\textsuperscript{v}{Institute for Astronomy, University of Hawaii, 2680 Woodlawn Drive, Honolulu, HI, 96822}\\
\textsuperscript{w}{Kapteyn Astronomical Institute, U. of Groningen, PO Box 800, 9700 AV Groningen, The Netherlands}\\
\textsuperscript{x}{Institut d'Astrophysique de Paris, UMR 7095, CNRS, 98 bis boulevard Arago, F-75014 Paris, France}\\
\textsuperscript{y}{Institut Lagrange de Paris, Sorbonne Universit\'{e}s, 98 bis Boulevard Arago, 75014 Paris, France}\\
\textsuperscript{z}{IPAC, Mail Code 314-6, California Institute of Technology, 1200 East California Boulevard, Pasadena, CA 91125, USA}\\ \textsuperscript{$\alpha$}{Homer L. Dodge Department of Physics \& Astronomy, University of Oklahoma, 440 W Brooks Street, Norman, OK 73019, USA}

\end{document}